\documentclass[a4paper,11pt]{article}
\pdfoutput=1 % if your are submitting a pdflatex (i.e. if you have
\usepackage{placeins,lineno}
\usepackage{jinstpub} % for details on the use of the package, please
\title{\boldmath A novel method for reconstruction of collision vertices at high-intensity hadron colliders}

\author[a,1]{F. Meloni\note{Corresponding author.}}

\affiliation[a]{University of Bern, Albert Einstein Center for Fundamental Physics, Laboratory for High Energy Physics
\\Sidlerstrasse 5,
\\3012 Bern,
\\Switzerland}

\emailAdd{federico.meloni@cern.ch}

\abstract{This paper presents a novel method for the reconstruction of interaction vertices in particle collision data. The algorithm is an agglomerative clustering technique designed for high-luminosity environments in current and future proton-proton colliders. Studies are presented in benchmark scenarios reproducing the LHC data-taking conditions and showing the performance as a function of the number of interactions per bunch crossing. Particular focus will be given to the foreseen data-taking conditions at the Run~3 of the LHC. The proposed algorithm is found to significantly improve the position resolution on the hard-scatter interaction by limiting the contamination of tracks from additional interactions in the vertex fit.}


\begin{document}
%\linenumbers
\maketitle
\flushbottom

\section{Introduction}
\label{sec:intro}

The study of the high-energy frontier in particle collisions has been, for years, the tool exploited for understanding the fundamental interactions that govern the Universe. The Large Hadron Collider (LHC)~\cite{Brüning:782076}, currently operating at CERN and delivering proton-proton ($pp$) collisions at a centre-of-mass energy $\sqrt{s}=13$~TeV, is the culmination of such research strategy. 
The next step is the increase of the intensity of the colliding proton beams, in order to access the rarest phenomena that appear in this energy regime. This is the main focus of ATLAS~\cite{PERF-2007-01} and CMS~\cite{Chatrchyan:2008aa} experiments in the High-Luminosity LHC (HL-LHC)~\cite{CERN-ACC-2014-0300} data-taking, and will be continued at Future Circular Collider facilities~\cite{Gomez-Ceballos:2013zzn,Acar:2016rde,Benedikt:2015poa} which will probe centre-of-mass energies of $\sqrt{s}=100$~TeV.

The estimation of the kinematic parameters and properties of particles produced in a collision event is of crucial importance in the analysis of the data in a high-energy physics experiment. The number of particles created by two colliding beams rises with the beam intensity. This is effect is particularly relevant for hadron colliders. 
The current average of simultaneous inelastic $pp$ interactions (pile-up) in a single proton bunch-crossing is about 20. It is expected to increase to about 80 during the 2021-2023 operation of the LHC (Run 3) and increase up to 300 during the HL-LHC (from 2026) data-taking. The average number of pile-up interactions under constant beam conditions will be denoted as $\mu$ in the following. These additional interactions, usually soft-QCD processes, will significantly increase the complexity of the topology of each recorded collision. 

A precise reconstruction of the primary vertex, defined as the point at which a $pp$ interaction occurred, directly affects the reconstruction of hard-scatter events. Since these events are the main object of study at the LHC, the correct assignment of the outgoing charged particle trajectories (tracks) to their primary vertex is essential in reconstructing the full kinematic parameters of a collision. This is of particular importance to identify the presence of non-interacting particles that escape detection, which is done by computing the measured momentum balance in the detector~\cite{Aad:2016nrq,ATL-PHYS-PUB-2015-027,Khachatryan:2014gga}. Vertex reconstruction~\cite{Aaboud:2016rmg,Chatrchyan:2014fea,Schluter:2014uea} allows to use an additional constraint to improve the momentum estimates of the particles involved in the event. Furthermore, the decay length of a short-lived particle can be determined by computing the distance between its estimated production and decay vertices~\cite{Aad:2015ydr,Chatrchyan:2012jua}.

The vertex finding algorithms have the role of clustering tracks into vertex candidates, which are then fitted to compute their final characteristics. The output of the vertex fit is a list of vertex positions as well as a set of updated track parameters for the particles associated to that particular production point.

This paper presents a feasibility study for a new method to reconstruct the primary $pp$ interaction vertices in collider experiments. Particular focus will be given to the foreseen conditions at the Run~3 of the LHC. Section~\ref{sec:algo} presents the effects arising from the presence of pile-up interactions and the proposed method. The performance will be presented in Section~\ref{sec:performance}, comparing with the algorithm currently in use in the ATLAS experiment at the LHC~\cite{ATL-PHYS-PUB-2015-026}.

\section{Algorithm description}
\label{sec:algo}

A vertex reconstruction algorithm must balance the efficiency for resolving close-by interactions in cases of large pile-up against the possibility of accidentally splitting a single, genuine interaction vertex into more than one cluster of tracks.

The performance of algorithms optimised for reconstructing a single interaction with the best possible precision will fail to perform in more densely populated environments: the contamination of tracks from nearby pile-up interactions in the vertex fit will progressively degrade the vertices spatial resolution. Furthermore, the presence of significant pile-up contamination can degrade the ability to identify the hard-scatter interaction among the reconstructed vertices. A typical selection criterion is based on the assumption that the charged particles produced in hard-scatter interactions have on average higher momentum than those produced in the pile-up collisions. However, the merging of multiple interactions can bias this choice, by selecting a pile-up vertex as hard-scatter.

The Global Vertex Clustering (GVC) algorithm is an agglomerative clustering technique, inspired by the methods used to reconstruct jets~\cite{Cacciari:2008}. A simultaneous reconstruction of the vertices of a collision is performed in order to allow local concentrations of tracks to form vertex candidates, even if close to a high-track-multiplicity interaction or located at the centre of the luminous region, i.e. the volume where collisions take place.
The algorithm is composed by the steps listed below.
\begin{itemize}
	\item A list of ``entities'' is created. These entities can be either single tracks or vertex candidates. Initially, an entity corresponding to each reconstructed track is included in the list.
	\item The set $d_{ij}$ of two-entity euclidean distances is computed. If two tracks are considered, $d_{ij}$ is the distance between the two points of closest approach to the beam axis. If a track and a vertex candidate are considered, $d_{ij}$ corresponds to the track's impact parameter, while if the vertex candidates are take into account, $d_{ij}$ is simply the distance between two points.
	\item If the two entities with lowest $d_{ij}$ satisfy a clustering rule, they are removed from the list. A new entity, grouping the tracks associated to the two entities, is added to the list. 
	\item The position of the new entity, now a vertex candidate, is recomputed by fitting a vertex from all the associated tracks.
	\item The procedure is repeated until no pair of entities satisfying the clustering rule is left.
\end{itemize}
In the following, the clustering rule is based on a threshold on the significance of the two-entity euclidean distance $\alpha = d_{ij}/\sigma_{d_{ij}}$, where $\sigma_{d_{ij}}$ is the uncertainty on $d_{ij}$. The ``perigee'' parametrisation~\cite{Billoir:1992yq} for helical tracks is used. The vertex position is fitted with a least-squares method~\cite{Piacquadio:2008zzb}. The GVC is a modular and parallelisable algorithm: the two main components, the clustering rule and the fitting strategy, can be developed independently.

\section{Performance}
\label{sec:performance}

In the ideal case of a fully efficient reconstruction, the number of reconstructed vertices scales linearly with $\mu$. In a realistic case, there are multiple effects that cause the relation to be non-linear: vertex merging, the splitting of a single interaction in multiple vertices and fake vertices arising from the combinatorial background. This section presents the performance of the GVC algorithm for a few relevant quantities, listed below.
\begin{itemize}
	\item The euclidean distance between pairs of adjacent vertices: for small distances, close-by interactions can no longer be resolved and are reconstructed as a single vertex. An algorithm is best when the minimal distance is determined by the detector intrinsic resolution.
	\item The hard-scatter track purity, defined as the ratio between the number of tracks coming from the hard-scatter and the total number of tracks associated to the reconstructed vertex: this is a direct measurement of the pile-up contamination in the vertex fit.
	\item The residuals of the vertex position, computed as the difference between the position at simulation level and its reconstructed position: the contamination of pile-up tracks in the vertex fit smears the resulting reconstructed position.
\end{itemize}

The vertex reconstruction performance has been estimated by ensemble testing with pseudo-experiments. A realistic distribution of the reconstructed tracks in the detector volume is simulated and used as inputs for the vertex reconstruction. The simulation neglects the vertex spatial distribution in the plane transversal to the beam direction. The position resolution in this plane is strongly constrained by the usage of the luminous region in the vertex fit~\cite{Aaboud:2016rmg} and leads to similar results irrespectively of the vertex reconstruction algorithm used. Interactions are generated along the beam axis ($z$) according to the shape of the luminous region observed in ATLAS in the 2015 data-taking~\cite{Beamspotpage}. This is modelled by a Gaussian distribution centred in $z=0$~mm, with a standard deviation of 45~mm. Events have been chosen to represent the reconstructed track distributions from one top quark pair production hard-scatter interaction and a variable number of pile-up interactions. A random track multiplicity is associated to each interaction according to a Poisson probability: the top quark pair interactions are represented by a group of tracks with an average multiplicity of 10, while the pile-up interactions by groups of tracks with an average multiplicity of 3~\cite{Aaboud:2016rmg}. Each track's $z$ position is distributed according to a Gaussian distribution with a standard deviation of 0.2~mm~\cite{Aaboud:2016rmg} centred on the real interaction position. Track reconstruction inefficiencies arising from  detector effects or acceptance are neglected. Contributions arising from tracks originating from decays of short-lived particles are neglected. 

In order to provide a realistic benchmark for the comparison, a private implementation of the vertex reconstruction algorithm adopted in the ATLAS experiment~\cite{Aaboud:2016rmg} has been used. It is an iterative procedure consisting of two major steps: seed finding and vertex fitting. The most probable position of the primary interaction is selected by looking for the maximum in the track multiplicity distribution along the beam axis. The vertex is fitted with an adaptive method and finally all the tracks found incompatible with the vertex by more than approximately 7 standard deviations are used to repeat the procedure until no more vertices with at least two associated tracks are found. The implementation of the ATLAS algorithm has been validated with a detailed comparison with the results presented in Refs.~\cite{Aaboud:2016rmg,ATL-PHYS-PUB-2015-026}. Good agreement is observed, with marginal differences attributed to the assumptions made on the track reconstruction efficiency and the position resolution used to determine the spatial distributions. 

\begin{figure}[htb!]
\begin{center}
\includegraphics[width=\textwidth]{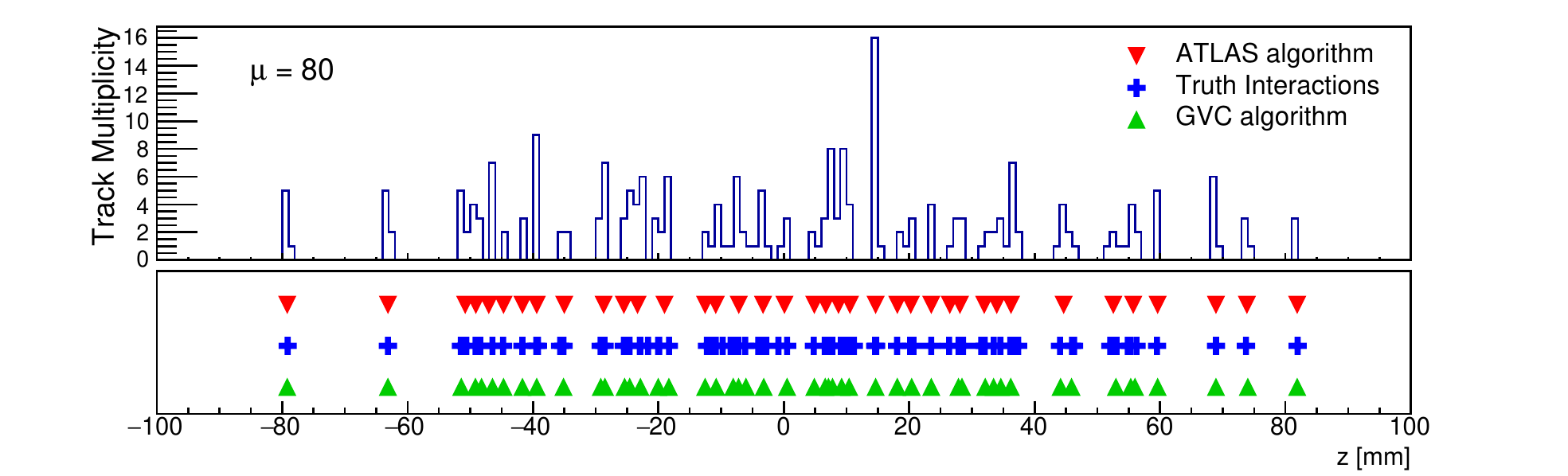}
\end{center}
\caption{Longitudinal track multiplicity distribution in a top quark pair production event, with 80 overlaid pile-up interactions, comparing the ATLAS algorithm (red triangles pointing down) with the GVC (green triangles pointing up). The blue crosses show the position of the interactions from the simulation truth information.} 
\label{fig:snapshot80}
\end{figure}

Each generated event is used as input for the vertex reconstruction algorithms under test, that are independently applied. 
Figure~\ref{fig:snapshot80} shows the output of the described simulation for a single event with 80 pile-up interactions. The observed track multiplicity is shown as a function of the position along the beam axis. The longitudinal distributions of the reconstructed vertices with the ATLAS and GVC algorithms are compared. In the central region (around $z = 0$~mm), where tracks are more closely distributed, it can be observed how the GVC algorithm is able to reconstruct more closely spaced interactions. 

\begin{figure}[htb!]
\begin{center}
\includegraphics[width=0.8\textwidth]{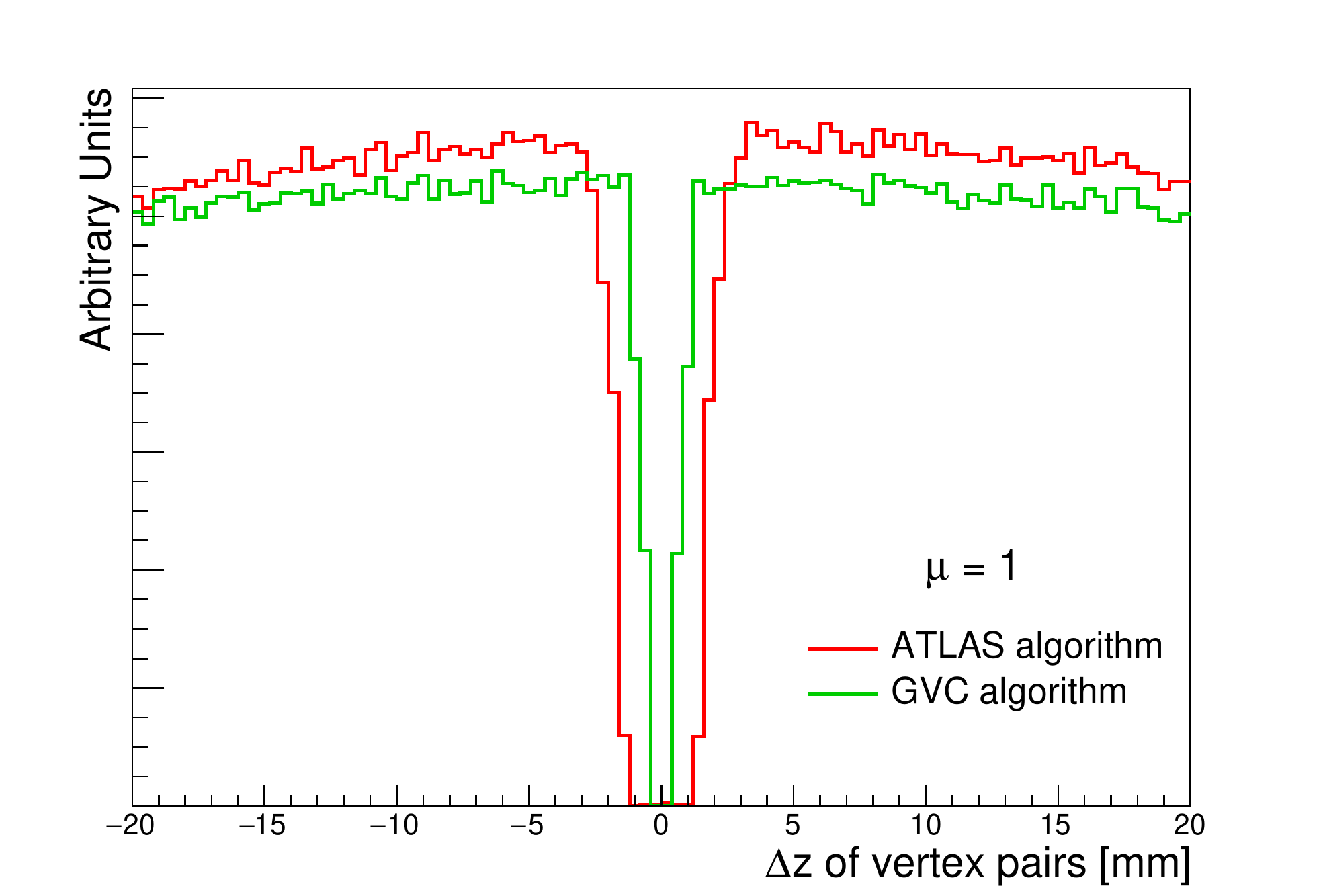}
\end{center}
\caption{Distribution of the longitudinal separation ($\Delta z$) between pairs of adjacent vertices, for soft-QCD events with $\mu=1$. The lines shows the expected distributions for the ATLAS (red) and GVC (green) algorithms.} 
\label{fig:deltaZ}
\end{figure}

The longitudinal separation $\Delta z$ between pairs of adjacent primary vertices is shown in Figure~\ref{fig:deltaZ}, considering events with only soft-QCD interactions and $\mu = 1$. This distribution is used to study the effects of vertex merging: the steep decrease of the number of reconstructed vertices at values of $\Delta z$ close to 0 is caused by this effect. The choice of a very low value of $\mu$ allows to disentangle the intrinsic resolution of the reconstruction algorithm from the presence of pile-up interactions. In the generated conditions, the GVC algorithm allows to resolve close-by interactions down to distances about a factor of three smaller than the ATLAS algorithm. A very small fraction of events reconstructed with the ATLAS algorithm contain vertices with a longitudinal separation below the typical primary vertex resolution: these are due to the splitting of a single interaction in multiple vertices. This effect does not apply to the GVC, since the clustering rule will group together all entities below the distance threshold defined in Section~\ref{sec:algo}.

\begin{figure}[htb!]
\begin{center}
\includegraphics[width=0.8\textwidth]{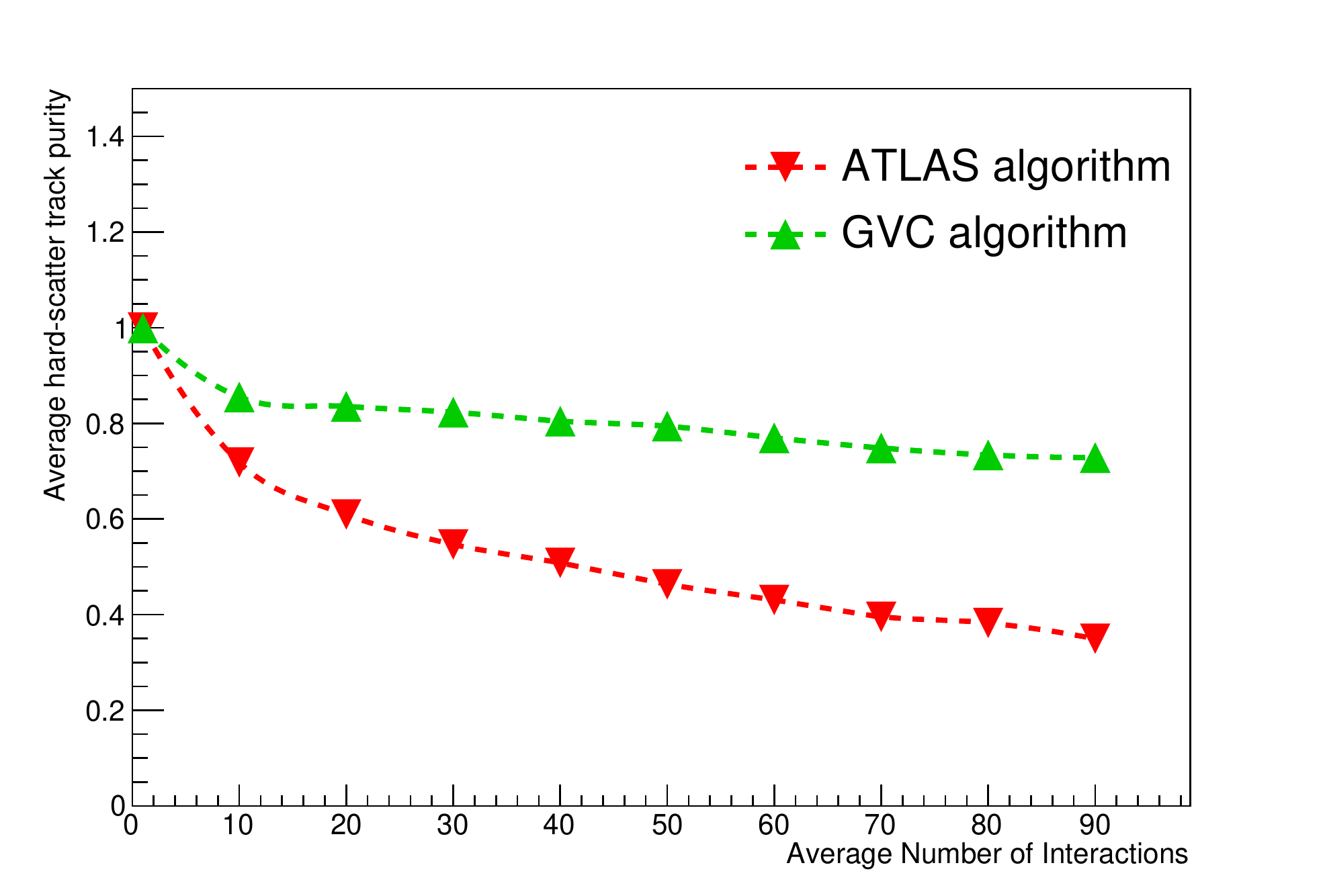}
\end{center}
\caption{Distribution of the average hard-scatter track purity as a function of the number of interactions, for the ATLAS (red triangles pointing down) and GVC (green triangles pointing up) algorithm.} 
\label{fig:PurityVsMu}
\end{figure}

The position of the hard-scatter interaction is the most important quantity to consider when using collision events for physics analysis. The average hard-scatter track purity as a function of $\mu$ is shown in Figure~\ref{fig:PurityVsMu}. 
Both algorithms have a significant contamination from pile-up tracks. However, the GVC has a consistently smaller pile-up track contamination with an improvement of about a factor of 2 in the $\mu$ range expected during the LHC Run 3.

\begin{figure}[htb!]
\begin{center}
\includegraphics[width=0.8\textwidth]{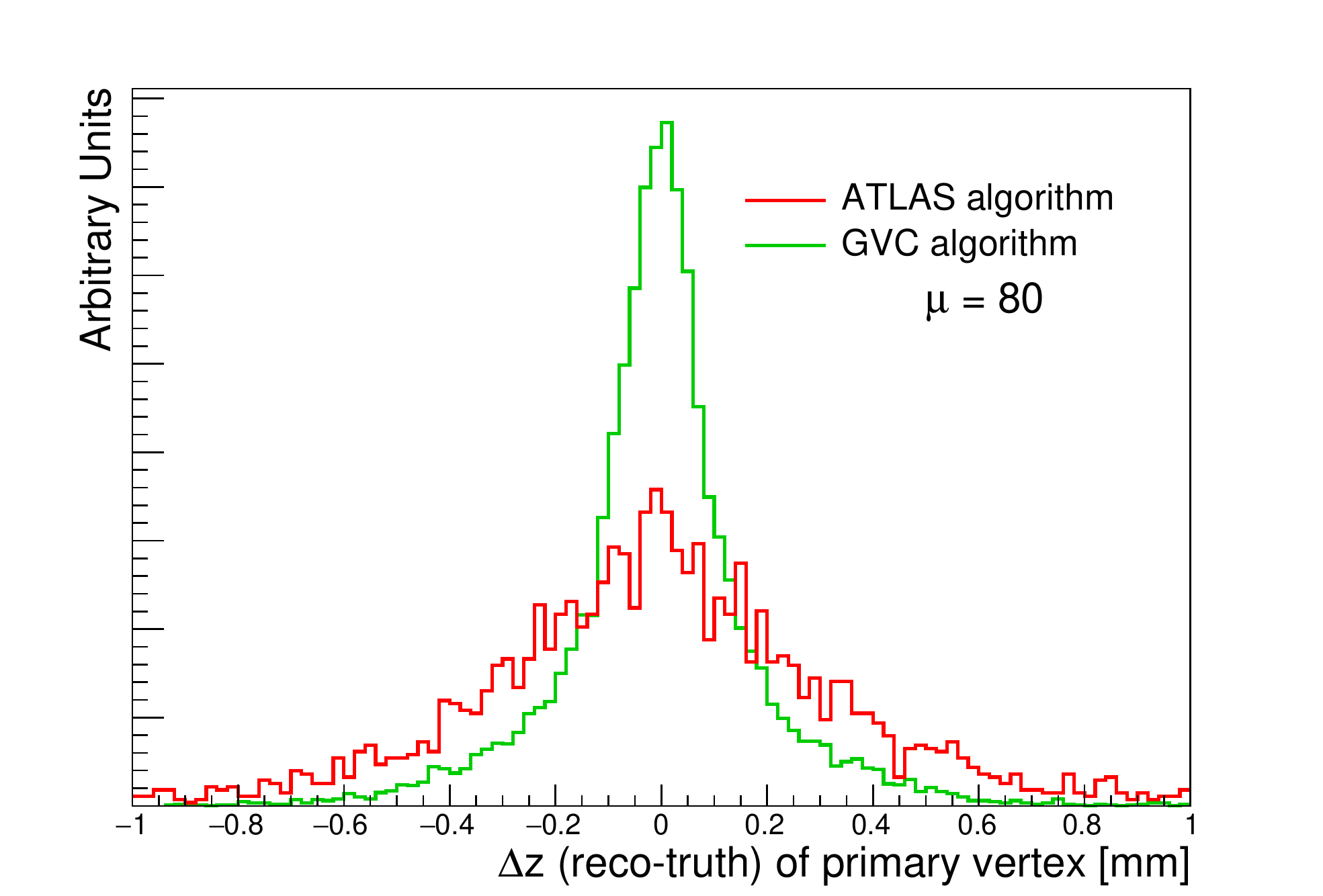}
\end{center}
\caption{The residual distributions in the $z$ coordinate for the reconstructed hard-scatter vertex in top pair production events with $\mu=80$. The performance of the ATLAS (red) and GVC (green) algorithms are shown by the histograms. The distributions are normalised to the same area.}
\label{fig:deltaZ_PV}
\end{figure}

This smaller pile-up contamination directly affects the distribution of residuals of the vertex position, shown in Figure~\ref{fig:deltaZ_PV}, and the vertex resolution. For events with $\mu=80$, the RMS of the residuals has been found to be of 0.18~mm for the GVC algorithm, while 0.33~mm for the ATLAS algorithm.

\begin{figure}[htb!]
\begin{center}
\includegraphics[width=0.8\textwidth]{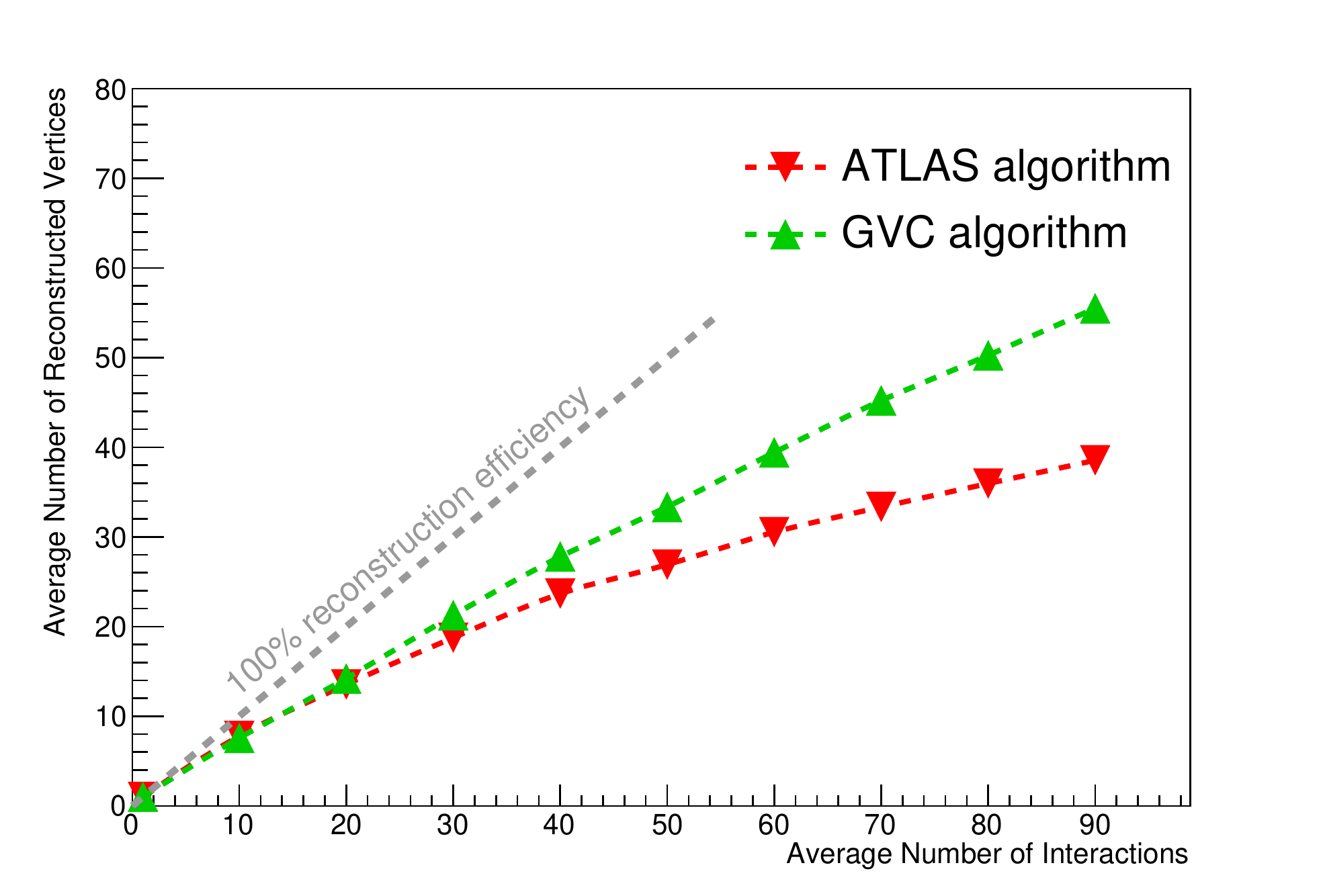}
\end{center}
\caption{Distribution of the average number of reconstructed vertices as a function of the number of interactions. The performance for the ATLAS (red triangles pointing down) and GVC (green triangles pointing up) algorithms are shown as dashed lines. The straight grey line represents the ideal case with perfect reconstruction efficiency.} 
\label{fig:AvgVxVsMu}
\end{figure}

The average number of reconstructed vertices by the two algorithms against the number of true interactions has also been studied, and is shown in Figure~\ref{fig:AvgVxVsMu}. This relation can be used to study the impact of the different reconstruction effects listed at the beginning of this Section. The merging of close-by interactions due to the detector and algorithmic resolution effects has the largest impact on primary vertex reconstruction efficiency as $\mu$ increases. The splitting of single interactions and reconstruction of fake vertices have been found to be negligible for both the algorithms taken into account. For a $\mu$ of about 80, the ATLAS algorithm has a reconstruction efficiency of about 40\%, while the GVC is closer to linearity with about 60\% of the interactions being correctly reconstructed. 

Finally, the average time needed by the GVC and ATLAS algorithms has been compared considering events with $\mu=80$ and has been found to be compatible within a few milliseconds.

\FloatBarrier

\section{Conclusions}
\label{sec:conclusions}

A proposal for a new vertex reconstruction algorithm has been made. A simplified simulation validated by reproducing the results published by the ATLAS experiment has been used to study the performances of such algorithm. The simulated data includes pile-up interaction multiplicities up to an average of 90, a value close to the one expected during the LHC Run 3 data-taking. 

Under these conditions, the GVC algorithm significantly improves the spatial resolution on the hard-scatter interaction by reducing the merging of close-by interactions. When considering an average number of pile-up interactions of 80, the hard scatter track purity is doubled. Nearby interactions are independently reconstructed down distances about a factor of three smaller than the ATLAS algorithm. The overall vertex reconstruction efficiency is increased by 16\%. The identification of short-lived particles and the computation of momentum balance in collider data will consequently be improved.

Future developments will target the optimisation of the algorithm for the expected beam conditions at the HL-LHC.

%\acknowledgments

%This is the most common positions for acknowledgments. A macro is available to maintain the same layout and spelling of the heading.

\bibliographystyle{BibStyle}
\bibliography{GVC_paper.bib}

\providecommand{\href}[2]{#2}\begingroup\raggedright\begin{thebibliography}{10}

\bibitem{Brüning:782076}
O.~S. Br{\"u}ning, P.~Collier, P.~Lebrun, S.~Myers, R.~Ostojic, J.~Poole, and
  P.~Proudlock, {\em {LHC Design Report}}.
\newblock CERN Yellow Reports: Monographs. CERN, Geneva, 2004.
\newblock \url{https://cds.cern.ch/record/782076}.

\bibitem{PERF-2007-01}
{ATLAS Collaboration}, {\em {The ATLAS Experiment at the CERN Large Hadron
  Collider}}, \href{http://dx.doi.org/10.1088/1748-0221/3/08/S08003}{JINST
  {\bfseries 3} (2008) S08003}.

\bibitem{Chatrchyan:2008aa}
{CMS} Collaboration, {\em {The CMS Experiment at the CERN LHC}},
\href{http://dx.doi.org/10.1088/1748-0221/3/08/S08004}{JINST {\bfseries 3}
  (2008) S08004}.
%%CITATION = JINST,3,S08004;%%.

\bibitem{CERN-ACC-2014-0300}
``{HL-LHC Preliminary Design Report: Deliverable: D1.5}.''
  {CERN-ACC-2014-0300}, 2015.
\newblock \url{https://cds.cern.ch/record/1972604}.

\bibitem{Gomez-Ceballos:2013zzn}
{TLEP Design Study Working Group}, {\em {First Look at the Physics Case of
  TLEP}}, \href{http://dx.doi.org/10.1007/JHEP01(2014)164}{JHEP {\bfseries 01}
  (2014) 164},
\href{http://arxiv.org/abs/1308.6176}{{\ttfamily arXiv:1308.6176 [hep-ex]}}.
%%CITATION = ARXIV:1308.6176;%%.

\bibitem{Acar:2016rde}
Y.~C. Acar, A.~N. Akay, S.~Beser, H.~Karadeniz, U.~Kaya, B.~B. Oner, and
  S.~Sultansoy, {\em {FCC Based Lepton-Hadron and Photon-Hadron Colliders:
  Luminosity and Physics}},
\href{http://arxiv.org/abs/1608.02190}{{\ttfamily arXiv:1608.02190
  [physics.acc-ph]}}.
%%CITATION = ARXIV:1608.02190;%%.

\bibitem{Benedikt:2015poa}
M.~Benedikt, B.~Goddard, D.~Schulte, F.~Zimmermann, and M.~J. Syphers, {\em
  {FCC-hh Hadron Collider - Parameter Scenarios and Staging Options}},
  {Proceedings, 6th International Particle Accelerator Conference (IPAC 2015):
  Richmond, Virginia, USA, May 3-8, 2015} (2015).
\url{http://accelconf.web.cern.ch/AccelConf/IPAC2015/papers/tupty062.pdf}.
%%CITATION = IPAC-2015-TUPTY062;%%.

\bibitem{Aad:2016nrq}
{ATLAS} Collaboration, {\em {Performance of algorithms that reconstruct missing
  transverse momentum in $\sqrt{s}=$ 8 TeV proton-proton collisions in the
  ATLAS detector}},
  \href{http://dx.doi.org/10.1140/epjc/s10052-017-4780-2}{Eur. Phys. J.
  {\bfseries C77} no.~4, (2017) 241},
\href{http://arxiv.org/abs/1609.09324}{{\ttfamily arXiv:1609.09324 [hep-ex]}}.
%%CITATION = ARXIV:1609.09324;%%.

\bibitem{ATL-PHYS-PUB-2015-027}
{ATLAS Collaboration}, ``{Performance of missing transverse momentum
  reconstruction with the ATLAS detector in the first proton--proton collisions
  at $\sqrt{s} = 13\;\mbox{TeV}$}.'' {ATL-PHYS-PUB-2015-027}, 2015.
\newblock \url{http://cdsweb.cern.ch/record/2037904}.

\bibitem{Khachatryan:2014gga}
{CMS} Collaboration, {\em {Performance of the CMS missing transverse momentum
  reconstruction in pp data at $\sqrt{s}$ = 8 TeV}},
  \href{http://dx.doi.org/10.1088/1748-0221/10/02/P02006}{JINST {\bfseries 10}
  no.~02, (2015) P02006},
\href{http://arxiv.org/abs/1411.0511}{{\ttfamily arXiv:1411.0511
  [physics.ins-det]}}.
%%CITATION = ARXIV:1411.0511;%%.

\bibitem{Aaboud:2016rmg}
{ATLAS} Collaboration, {\em {Reconstruction of primary vertices at the ATLAS
  experiment in Run 1 proton-proton collisions at the LHC}},
\href{http://arxiv.org/abs/1611.10235}{{\ttfamily arXiv:1611.10235
  [physics.ins-det]}}.
%%CITATION = ARXIV:1611.10235;%%.

\bibitem{Chatrchyan:2014fea}
{CMS} Collaboration, {\em {Description and performance of track and
  primary-vertex reconstruction with the CMS tracker}},
  \href{http://dx.doi.org/10.1088/1748-0221/9/10/P10009}{JINST {\bfseries 9}
  no.~10, (2014) P10009},
\href{http://arxiv.org/abs/1405.6569}{{\ttfamily arXiv:1405.6569
  [physics.ins-det]}}.
%%CITATION = ARXIV:1405.6569;%%.

\bibitem{Schluter:2014uea}
{Belle-II Software Group} Collaboration, {\em {Vertexing and Tracking Software
  at Belle II}}, PoS {\bfseries Vertex2014} (2014) 039,
\href{http://arxiv.org/abs/1411.3485}{{\ttfamily arXiv:1411.3485
  [physics.ins-det]}}.
%%CITATION = ARXIV:1411.3485;%%.

\bibitem{Aad:2015ydr}
{ATLAS} Collaboration, {\em {Performance of $b$-Jet Identification in the ATLAS
  Experiment}}, \href{http://dx.doi.org/10.1088/1748-0221/11/04/P04008}{JINST
  {\bfseries 11} no.~04, (2016) P04008},
\href{http://arxiv.org/abs/1512.01094}{{\ttfamily arXiv:1512.01094 [hep-ex]}}.
%%CITATION = ARXIV:1512.01094;%%.

\bibitem{Chatrchyan:2012jua}
{CMS} Collaboration, {\em {Identification of b-quark jets with the CMS
  experiment}}, \href{http://dx.doi.org/10.1088/1748-0221/8/04/P04013}{JINST
  {\bfseries 8} (2013) P04013},
\href{http://arxiv.org/abs/1211.4462}{{\ttfamily arXiv:1211.4462 [hep-ex]}}.
%%CITATION = ARXIV:1211.4462;%%.

\bibitem{ATL-PHYS-PUB-2015-026}
{ATLAS Collaboration}, ``{Vertex Reconstruction Performance of the ATLAS
  Detector at$\sqrt{s} = 13\;\mbox{TeV}$}.'' {ATL-PHYS-PUB-2015-026}, 2015.
\newblock \url{http://cdsweb.cern.ch/record/2037717}.

\bibitem{Cacciari:2008}
M.~Cacciari, G.~P. Salam, and G.~Soyez, {\em {The anti-kt jet clustering
  algorithm}}, \href{http://dx.doi.org/10.1088/1126-6708/2008/04/063}{JHEP
  {\bfseries 04} (2008) 063}, \href{http://arxiv.org/abs/0802.1189}{{\ttfamily
  arXiv:0802.1189 [hep-ph]}}.

\bibitem{Billoir:1992yq}
P.~Billoir and S.~Qian, {\em {Fast vertex fitting with a local parametrization
  of tracks}},
\href{http://dx.doi.org/10.1016/0168-9002(92)90859-3}{Nucl. Instrum. Meth.
  {\bfseries A311} (1992) 139--150}.
%%CITATION = NUIMA,A311,139;%%.

\bibitem{Piacquadio:2008zzb}
G.~Piacquadio, K.~Prokofiev, and A.~Wildauer, {\em {Primary vertex
  reconstruction in the ATLAS experiment at LHC}},
\href{http://dx.doi.org/10.1088/1742-6596/119/3/032033}{J. Phys. Conf. Ser.
  {\bfseries 119} (2008) 032033}.
%%CITATION = 00462,119,032033;%%.

\bibitem{Beamspotpage}
{ATLAS Collaboration}, {\em {ATLAS Beamspot Public Results}}, 2015.
\newblock
  \url{https://twiki.cern.ch/twiki/bin/view/AtlasPublic/BeamSpotPublicResults}.

\end{thebibliography}\endgroup

\end{document}